\documentclass[twocolumn,showpacs,prb,10pt]{revtex4}
\usepackage[english]{babel}
\usepackage{amsmath,amssymb}
\usepackage{times}
\usepackage{epsfig}

\begin{document} 
\title{Luttinger sum rule for finite systems of correlated electrons}

\author{J. Kokalj$^{1}$ and P. Prelov\v sek$^{1,2}$}
\affiliation{$^1$J.\ Stefan Institute, SI-1000 Ljubljana, Slovenia}
\affiliation{$^2$ Faculty of Mathematics and Physics, University of
Ljubljana, SI-1000 Ljubljana, Slovenia}

\date{\today}
                   
\begin{abstract}
The validity of the Luttinger sum rule is considered for finite
systems of interacting electrons, where the Fermi volume is determined
by location of zeroes of Green's function. It is shown that the sum
rule in the paramagnetic state is evidently violated within the planar
$t$-$J$ model at low doping while for the related Hubbard model, even
in the presence of next-nearest-neighbor hopping, no clearcut
exception is found.
\end{abstract}

\maketitle

\section{Introduction}

In last decades extensive experimental and theoretical investigations
of high-temperature superconductors and other novel materials with
strongly correlated electrons raised several fundamental questions
regarding the Fermi liquid (FL) concept. One of the central
assumptions in the FL theory is the existence of well defined Fermi
surface with the volume, independent of electron-electron
interactions. The latter property has firm basis in the Luttinger sum
rule (LSR) \cite{lutt1,lutt2,abri}, which represents the relation
between the electron density and the poles or zeroes of the
single-particle Green's function (GF) $G({\bf k}, \omega)$ at the
chemical potential.

To settle this issue several authors reconsidered the LSR, its validity
and extensions regarding different aspects in connection with strongly
interacting electrons. It has been shown that in one dimension (1D)
LSR remains robust, although the usual FL is replaced by the
Luttinger-liquid phenomenology \cite{blag}. LSR is valid even for
models of strongly correlated electrons as the Kondo lattice model
\cite{yama}. Recently, the LSR in the Mott insulating state has
attracted attention. In this case the chemical potential is in the gap
so there is no Fermi surface, however the LSR can still apply for
zeroes of GF and in this way the definition of Fermi surface is
generalized to the concept of 'Luttinger' surface \cite{dzya} and
corresponding Luttinger volume (LV). In the insulator the LSR should
remain valid for models with the electron-hole symmetry as the
Hubbard model with the nearest-neighbor hopping on a bipartite lattice
\cite{stan}. On the other hand, extensions of the latter model can
lead to the violation of the LSR in the insulating state \cite{rosc},
or a novel form in ladder systems \cite{koni}. The LSR has to be also
reformulated in the case of emergence of long range order \cite{alts}.
Clearly, the most challenging question is the metallic state of
strongly correlated electrons in the absence of any broken symmetry.
There are indications from photoemission experiments that at least on
some hole-doped cuprates the Luttinger volume is not conserved
\cite{shen,yosh}.  Similar conclusions can be drawn from numerical
studies of related relevant models as the $t$-$J$ model \cite{putt}
and Hubbard model \cite{maie} near the Mott insulating state although
deviations are modest.

In this paper we consider the LSR on finite systems of correlated
electrons. We use the observation that the main ingredients of the
theorem \cite{lutt1} are valid for finite systems, exploited rarely so
far \cite{praz}. The underlying idea is that in this way one can
directly test nontrivial models in different parameter regimes on the
validity of LSR and the topology of the related Fermi surface and
LV. While in this way it is still hard to prove the Luttinger theorem
in the most relevant thermodynamic limit, it is much easier to show on
its breakdown. As shown later, one should be careful in the
classification of scenarios of LSR violation, since only some of them
could remain relevant up to the thermodynamic limit.

We investigate as examples two prototype models of strongly correlated
electrons, most frequently studied in connection with superconducting
cuprates, namely the two-dimensional (2D) Hubbard model and the
$t$-$J$ model on a square lattice. For comparison, we perform tests
also for corresponding 1D models whereby the 1D Hubbard model is
exactly solvable \cite{lieb,frah} and the LSR should remain valid in the
thermodynamic limit. Within our numerical limitations presented
results show that also 2D Hubbard model generally fulfills the LSR in
the whole parameter range of the paramagnetic state, even in the
presence of the next-nearest-neighbor hopping $t'$. Still, we find
some exception which we classify as not clearcut. On the other hand,
the 2D (and even 1D) $t$-$J$ model reveals evident deviations from LSR
close to half filling.

The paper is organized as follows. In Sec.~II we briefly summarize the
formalism underlying the Luttinger sum rule, in particular its
relevance and application for finite-size systems. We introduce the
tight-binding models for interacting electrons, whereby the Hubbard
model is the simplest one allowing the study of continuous
development from the non-interacting case satisfying trivially LSR
into the strong-correlation regime. The approach of avoiding the
degeneracy with twisted boundary condition is described. We also
clasify several possible scenarios of the LSR breakdown in finite
systems. In Sec.~III we present numerical tests of the LSR for the
Hubbard model and the $t$-$J$ model, both on 1D chains and 2D square
lattice. In particular, we try to pinpoint clearcut cases of the
LSR breakdown, which are evident for 2D $t$-$J$ model near half-filling.
Sec.~IV is devoted to conclusions and discussion.
  
\section{Luttinger sum rule}

\subsection{Formalism for finite systems}

We consider here the homogeneous system of interacting electrons on a
lattice with periodic boundary conditions. In this case the LSR
\cite{lutt1,lutt2,abri} can be expressed as
\begin{equation}
N=\sum_{\mathbf{k}s, G_{s}(\mathbf{k},0) >0} 1, \label{eq1}\end{equation}
relating the number of fermions $N$ to the value of zero-temperature
$T=0$ single-particle GF $G_{s}(\mathbf{k},\omega=0)$ at the chemical
potential $\mu$. General $\mathrm{Tr}$ \cite{lutt1} is in homogeneous
case replaced by the sum over ${\bf k}$ and spin $s$.

Let us recall few essential steps in the derivation of Eq.~(\ref{eq1})
\cite{lutt1,abri} in order to see possible limitations and its proper 
application for finite systems. Generally one can express
\begin{equation}
N= \frac{1}{\beta} \sum_l  \sum_{\mathbf{k}s}
{\cal G}_{s}(\mathbf{k},\omega_l)   e^{i \omega_l 0^+},
\label{eq2}
\end{equation}
where ${\cal G}_{s}(\mathbf{k},\omega_l)$ is the $T>0$ propagator
at Matsubara $\omega_l=2\pi(2l+1)/\beta$ and $\beta=1/T$ (we use units
$\hbar = k_B=1$). With the definition of the self energy
$\Sigma_{s}(\mathbf{k},\zeta)=\zeta -\epsilon_s(\mathbf{k})
-1/{\cal G}_{s}(\mathbf{k}, \zeta)$ Eq.~(\ref{eq2}) can be rewritten
as $N=I_1+I_2$ \cite{lutt1,abri},
\begin{eqnarray}
&&I_1= -\frac{1}{2 \pi i} \sum_{\mathbf{k}s} \int_{\Gamma} 
\frac{\partial}{\partial \zeta} \ln ({\cal G}_s
(\mathbf{k},\zeta)) e^{\zeta 0^+} 
\frac{1}{e^{\beta \hbar \zeta}+1} d\zeta \\
&&I_2= \frac{1}{2 \pi i} \sum_{\mathbf{k}s} \int_{\Gamma} 
{\cal G}_s(\mathbf{k},\zeta)
\frac{\partial}{\partial \zeta}\Sigma_s(\mathbf{k}, \zeta)
e^{\zeta 0^+} \frac{1}{e^{\beta \hbar \zeta}+1} d\zeta, \nonumber
\label{eq3}
\end{eqnarray}
where integration path $\Gamma$ clockwise encloses real axis.  The
central point of the proof of the LSR is the observation that $I_2$ in
Eq.~(\ref{eq3}) vanishes after limiting $T\to 0$. The latter is argued
with the construction of the functional $Y'$, which is the
contribution of all closed linked skeleton diagrams to thermodynamic
potential. $I_2$ can be represented as a full derivative of $Y'$,
hence $I_2= \partial Y'/\partial \delta \epsilon=0$
\cite{lutt1,abri}. We note that the 
existence and convergence of $Y'$ is shown within the perturbation
theory, which applies also for finite systems. It should be pointed
out that deviations from LSR in concrete case discussed later on can be
traced back to the nonvanishing $I_2$ and in this way to the breakdown
of perturbation theory for $Y'$.

In regular cases we have $N=I_1$ which for $T \to 0$ reduces to
\begin{equation}
N=-\frac{1}{2 \pi i} \sum_{\mathbf{k}s} 2 \textrm{Im} \big{[}
\int_{-\infty}^{ 0} d\zeta \big{\{} \frac{\partial}{\partial \zeta}
\ln (G_{s}(\mathbf{k},\zeta)) \big{\}}
\big{]}. \label{eq4}
\end{equation}
In finite systems we can express the GF for $N$ electrons explicitly
in terms of eigenstates of systems with $N-1,N, N+1$ particles,
\begin{eqnarray}
G_{s}(\mathbf{k},\zeta)&=& \sum_{m}
\frac{\big{|} \langle m_{N-1}|c_{\mathbf{k}s} |0_N \rangle
\big{|}^2} {\zeta+ \mu_N -(E_0^N-E_m^{N-1})} + \nonumber \\
&+& \sum_{l} \frac{\big{|}
\langle l_{N+1}|c^\dagger_{\mathbf{k}s} |0_N \rangle
\big{|}^2} {\zeta+ \mu_N -(E_l^{N+1}-E_0^{N})}. \label{eq5}
\end{eqnarray}
$\mu_N$ in Eq.~(\ref{eq5}) is according to derivation \cite{lutt1}
defined by the grand-canonical relation $N=T
\partial(\ln \Omega)/\partial \mu$ in the limit $T \to 0$.  The latter
gives uniquely $\mu_N = (E_0^{N+1} - E_0^{N-1})/2$ under the provision
that $E_0^N$ is a concave function of $N$ or at least $E_0^{N+1} +
E_0^{N-1} - 2 E_0^{N} > 0$, which is equivalent to the phase stability
of the ground state (g.s.). It should be noted that in finite systems
the latter condition is not fulfilled in certain case which can be
signature of a physical instability but as well just a finite-size
effect. Another source of ambiguities in finite size systems can arise
from the degeneracy of the ground state $|0_N\rangle$, which can be
partly removed by introduction of appropriate fields and twisted
boundary conditions.

It is easy to see that the contributions to Eq.~(\ref{eq4}) come from
poles and zeros of $G_{s}(\mathbf{k},\omega<0)$. The latter appear in
pairs, and give residuums $\pm 1$, respectively, mostly cancelling
each other. As a rule we establish that finally the LSR,
Eq.~(\ref{eq1}), is determined by the unpaired zero of GF at $\omega
\sim 0$ which can appear either for $\omega \geq 0$ or $\omega \leq
0$. This is a distinction to a macroscopic systems where the Fermi
surface is located by the poles of GF at $\omega =0$ . On the other
hand, in a finite system the poles of GF are at $\omega \ne 0$, since
in Eq.~(\ref{eq5}) in general $E_0^N-E_0^{N-1}<\mu_N$ and
$E_0^{N+1}-E_0^N>\mu_N$ due the choice of $\mu_N$ and concavity of
$E_0^N$. Clearly, possible exceptions are when the latter condition is
not satisfied as well when the g.s. is degenerate.

\subsection{Tight binding models}

In the following we test the validity of LSR for single-band
tight-binding models for correlated electrons,
\begin{equation}
H = - \sum_{i,j,s} t_{ij} c_{js}^\dagger c_{is}+ H_{int},\label{eq6}
\end{equation}
where the prototype model is the Hubbard model with local repulsion
\begin{equation}
H_{int}=U \sum_i n_{i \uparrow} n_{i\downarrow}, \label{eq6a}
\end{equation}
and the second model is the $t$-$J$ model discussed lateron.  We
consider finite systems with $N_0$ sites on 1D chain and 2D square
lattice. Besides the nearest-neighbor hopping $t_{ij}=t$ we
investigate also the next-nearest-neighbor hopping $t_{ij}=t'$, since
the latter breaks the electron-hole symmetry, e.g. in the Hubbard
model, which may affect the LSR \cite{stan}. The Hubbard model has the
advantage that by increasing $U$ one can study continuous development
from the reference system of noninteracting electrons (NIE) on a
lattice, where the LSR is trivially satisfied, into a strong
correlation regime where the breakdown of LSR can appear in several
ways. It is evident that the interaction $U$ can be treated within the
standard perturbation theory and therefore the arguments underlying
the proof of LSR \cite{lutt1} should apply.

We note that in finite systems even for $U=0$ a degeneracy of the
g.s. can appear in the case of periodic boundary conditions due to
discrete single-particle ${\bf k}={\bf k}_i, i=1,N_0$. This can be
removed by introducing the twisted boundary conditions which are
achieved by replacing $t_{ij} \to t_{ij} \exp(i \theta {\bf r}_{ij})$
while the interaction in Eq.~(\ref{eq6}) is supposed not to depend on
$\theta$. In this case, we are dealing on 2D square lattice with the
reference system of NIE with the single-particle dispersion
\begin{equation}
\epsilon({\bf k})=-2t(\cos \tilde k_x+ 
\cos \tilde k_y)-4t'\cos \tilde k_x \cos \tilde k_y, \label{eq7}
\end{equation} 
where $\tilde {\bf k}={\bf k}+{\bf \theta}$.  To remove degeneracies,
it is enough to take $\theta=(\theta_x,\theta_y)$ infinitesimally
small, although also finite $\theta$ may have a meaning, e.g. by
minimizing the g.s. energy $E_0^N(\theta)$.

\subsection{Scenarios of sum-rule violation} 

Before presenting results of our analysis, let us classify possible
ways of the LSR violation. We note that generally eigenstates of
Eq.(\ref{eq6}) can be sorted with respect to several quantum numbers:
number of electrons $N$, total spin projection $S^z$ (as well as $S$),
total momentum ${\bf K}$, etc. According to the latter several
scenarios are possible, e.g., for the Hubbard model, bearing in mind
that the reference NIE system is paramagnetic:
 
\noindent I) Turning on $U>0$ the character of the g.s. $|0_N\rangle$
can change due to the crossing of levels with different
symmetries. This is manifested by an abrupt jump of $G({\bf k},0)$ and
possible violation of LSR. Such a case can be a signature of a
macroscopic phase change, e.g., for $S \gg 0$ \cite{alts} indicating a
ferromagnetic instability (e.g., for $U \to \infty$ and $N=N_0 \pm 1$
well known Nagaoka instability), or merely a finite size effect
(mostly change of ${\bf K}$).

\noindent II) Similar, but more subtle, could be the effect of the
level crossing in $|0_{N+1}\rangle$ or $|0_{N-1}\rangle$.  In this
case, $\mu_N$ is determined by the g.s. with $N+1,N-1$ electrons,
while $G({\bf k},0)$, Eq.~(\ref{eq5}) does not necessarily contain
matrix elements between $N-1,N,N+1$ g.s., if e.g. $|S_{N+1}-S_N|>
1/2$. As in scenario I such a level crossing could be a sign of a
macroscopic instability.
 
\noindent III) The g.s. of NIE can be degenerate on finite lattices. The
latter can be eliminated by introducing small $\theta$, but is removed
also by $U>0$. Such a case with possible level crossing
between different ${\bf K}$ at small $U^*(\theta)$ could lead to
ambiguity, but is not the problem if the breakdown of LSR appears at
larger $U_c \gg U^*$.

\noindent IV) To avoid degeneracies of NIE, most rewarding are
configurations with closed shells of electrons where we can fix
$\theta=0$. Then most clearcut (excluding scenarios I, II) breakdown
of the LSR, Eq.~(\ref{eq1}), can appear when by increasing $U>U_c$ one
or several zeroes of $G({\bf k},\omega_0)$ cross the chemical
potential, i.e. $\omega_0$ changes sign.

\section{Numerical results}

In the following we present numerical results obtained for the 1D and
2D Hubbard model and $t$-$J$ model. In 2D lattices are chosen of
the Pythagorean form $N_0=L^2+M^2$. To avoid complications in the
interpretation we consider only cases with even $N, N_0$. For smaller
sizes, $N_0 \leq 8$ one can evaluate $G_s({\bf k},\omega)$ via
Eq.~(\ref{eq5}) finding all eigenstates using the exact (full)
diagonalization within the basis states for given $N,S^z$. To reach
larger sizes, i.e. $N_0=16$ for Hubbard model and $N_0=20$ for the
$t$-$J$ model, respectively, we calculate the g.s. performing the
exact diagonalization with the Lanczos technique, which is then also
applied to calculate $T=0$ $G_s({\bf k},\omega)$ in a usual way
\cite{dago}. It can be shown that such a procedure yields acute
values for quantities of interest, in particular $E_0^N, \mu_N$ and
$G_s({\bf k},0)$.

\subsection{Hubbard model}

\noindent{\bf 1D systems}: For the start we test the basic $t'=0$ 
Hubbard model, Eq. (\ref{eq6}), on a chain with up to $N_0=14$
sites. As expected we find no violation of the LSR in cases with even
$N$ in the investigated range of $0<U<30~t$.  Adding $t'\neq 0$ breaks
the particle-hole symmetry \cite{stan} and even at half-filling LSR
could become questionable \cite{rosc}. Nevertheless, choosing
$t'=-0.2~t$ we also do not find any evident breakdown following
scenario IV of LSR at or away from half-filling.

Several regimes with the violation of LSR (e.g., $N=6$ electrons on
$N_0=14$ sites for $U>6~t$)  can be mainly
attributed to level crossing of $|0_{N+1}\rangle$ or $|0_{N-1}\rangle$
(scenario II). Also for half filled $N_0 = 12$ system, where the
degeneracy of the g.s. is removed by choosing small $\theta$ we find
$N'=10$ particles in LV instead of $N=12$ above $U_c(\theta)<4~t $,
which seems to represent the violation of type III.  In this case
$N-1,N,N+1$ g.s. preserve symmetry of NIE, however $|0_N\rangle$ has
nonzero ${\bf K}$.

\noindent{\bf 2D systems}: $t'=0$ Hubbard model we 
investigate on square lattices with $N_0=8,10,16$ sites.  As a general
rule (with some exceptions elaborated below), we find that the LSR
remains satisfied for all considered $N \leq N_0$ and moreover the
topology of the LV remains that of NIE.  In particular, we find no
violation for half filling $N=N_0$, consistent with particle-hole
symmetry \cite{stan,rosc}.  

Away from $N=N_0$ one exceptional deviating case is $N=6$ on $N_0=8$
sites where at larger $U>U_c$ the LSR yields $N'=8,10,14>N$. Again, in
this case NIE g.s. is degenerate and $U_c$ is $\theta$ dependent and
also for $U<U_c$ level crossing in $N-1,N+1$ g.s. is observed. All
this indicates on combination of scenarios II and III.

In our study of 2D $t$-$t'$-$U$ model we use systems with
$N_0=8,10,16$ sites and $t'=-0.3t$ as frequently invoked for
superconducting cuprates. We find no clear (type IV) violation of LSR
in range $0<U<40~t$ and different fillings $N \leq N_0$. However, some
deviations from LSR connected with level crossing scenario II are
found at and away form half-filling even for closed shell
configurations. E.g., $N=14$ electrons on $N_0=16$ sites at $U>30~t$
result in $N'=20$ within the LV. Although $|0_N\rangle$ g.s. remains
throughout $S=0$ and $K=0$, in this case $N+1$ g.s. changes from NIE
value $S=1/2$ to higher spin $S>1/2$ at $U>30~t$.

\subsection{$t$-$J$ model}

Another prototype model for strongly correlated electrons is the
$t$-$J$ model,
\begin{equation}
H = - \sum_{i,j,s}t_{ij} \tilde{c}_{js}^\dagger  \tilde{c}_{is}
+ J \sum_{<ij>} (\mathbf{S}_i \cdot \mathbf{S}_j - \frac{1}{4} n_i n_j ),
\label{eq8}
\end{equation}
where $\tilde{c}_{is} = c_{is}(1-n_{i,-s})$ are projected operators
not allowing for a double occupancy at each site, and $\mathbf{S}_i$
are the local spin operators coupled with the exchange interaction
$J$. The $t$-$J$ model can be considered as the truncated Hubbard
model at large $U\gg t$ and $J=4t^2/U$. It is assumed that low-energy
physics of both models is similar, therefore it is of interest whether
they behave similarly with respect to LSR. We note that formally
$t$-$J$ model can be also written as the tight-binding model,
Eq.~(\ref{eq6}), with the interaction
\begin{equation}
H_{int}= J \sum_{<ij>} (\mathbf{S}_i \cdot \mathbf{S}_j - \frac{1}{4}
n_i n_j )+\tilde U \sum_i n_{i\uparrow} n_{i\downarrow}, \label{eq8a}
\end{equation}
whereby we let $\tilde U \to \infty$ to get a projected model,
Eq.~(\ref{eq8}). Such a formulation helps to formally connect the
$t$-$J$ model with finite $J>0$ continuously with NIE. Although the
model is not perturbative in any limit, one can first consider the
$J>0$ term as the perturbation and then performing the limit $\tilde
U\to \infty $. In this way it makes sense to compare the LSR within
the $t$-$J$ model directly to NIE with the dispersion,
Eq.({\ref{eq7}).

\noindent {\bf 1D systems:} Within the $t$-$J$ model it makes sense to
discuss cases $N<N_0$, i.e. representing the Mott insulator doped with
$N_h=N_0-N$ holes. The model on a chain at $J\to 0$ behaves as the
$U=\infty$ Hubbard model obeying LSR as confirmed in our numerical
calculations. Finite $J>0$ is nontrivial and some breakdowns of LSR
are found at low doping $N_h \ll N_0$. For $N_0=16$ and $N_h=2, N=14$
representing a closed shell configuration we find at $J>0.7~t$
$N'=24$. Since the $|0_N\rangle$ has $S=0, K=0$ as well as $N-1$,$N+1$
g.s. have $S=1/2$, this case represents a clean violation of LSR of
type IV.

\noindent{\bf 2D systems:} Here, one should also monitor the concavity 
of $E^0_N$ which can be violated at low doping, but not for even $N$
considered in our study. In general, results and conclusions for the
LV within the $t$-$J$ model are quite similar to the corresponding
large-$U$ Hubbard model. Essential differences appear at higher
parameter $J>0.1~t$ and at $N_h \ll N_0$.
We investigate further on lattices with $N_0=16, 18, 20$ sites
and $N_h<6$.

The most clear counterexample of LSR following scenario IV is the
closed-shell system with $N_h=2$ on $N_0=20$ sites. The corresponding
LV for NIE is shown Fig.~1a, together with wave-vectors appearing on a
$N_0=20$ lattice. On the other hand, we find that for $0.1~t <J <
0.3~t$ the LV contains $N'=22$ particles, as shown in Fig.~1b. Here,
$|0_N \rangle$ has $S=0, K=0$, and no level crossing in $N-1,N+1$
g.s. with $S=1/2$ is detected in this regime, excluding scenario II. 
For larger $J/t$ even more drastic violations are found. For $J=0.4~t,
J=0.5~t$ we find $N'=24$ and $N'=32$, respectively, as represented in
Figs.1c,d. It is expected that introduction of $t'$ should even
increase the deviation from the LSR \cite{zeml}.

\begin{figure}
\begin{minipage}[c]{0.22 \textwidth}
  a) NIE
  \includegraphics[width =\textwidth] {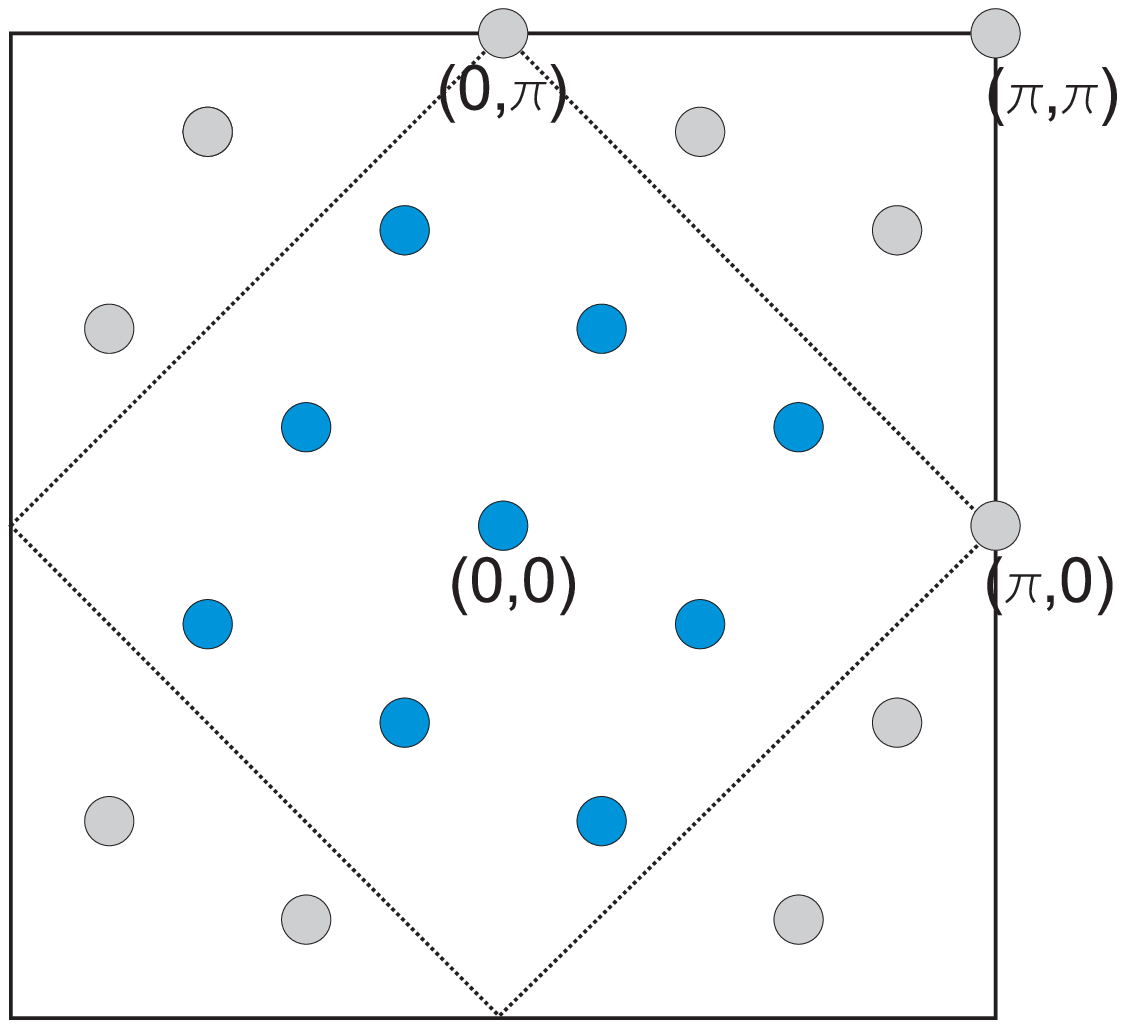}  
\end{minipage}
\begin{minipage}[c]{0.22 \textwidth}
  b) $0.1~t<J<0.3~t$
  \includegraphics[width =\textwidth] {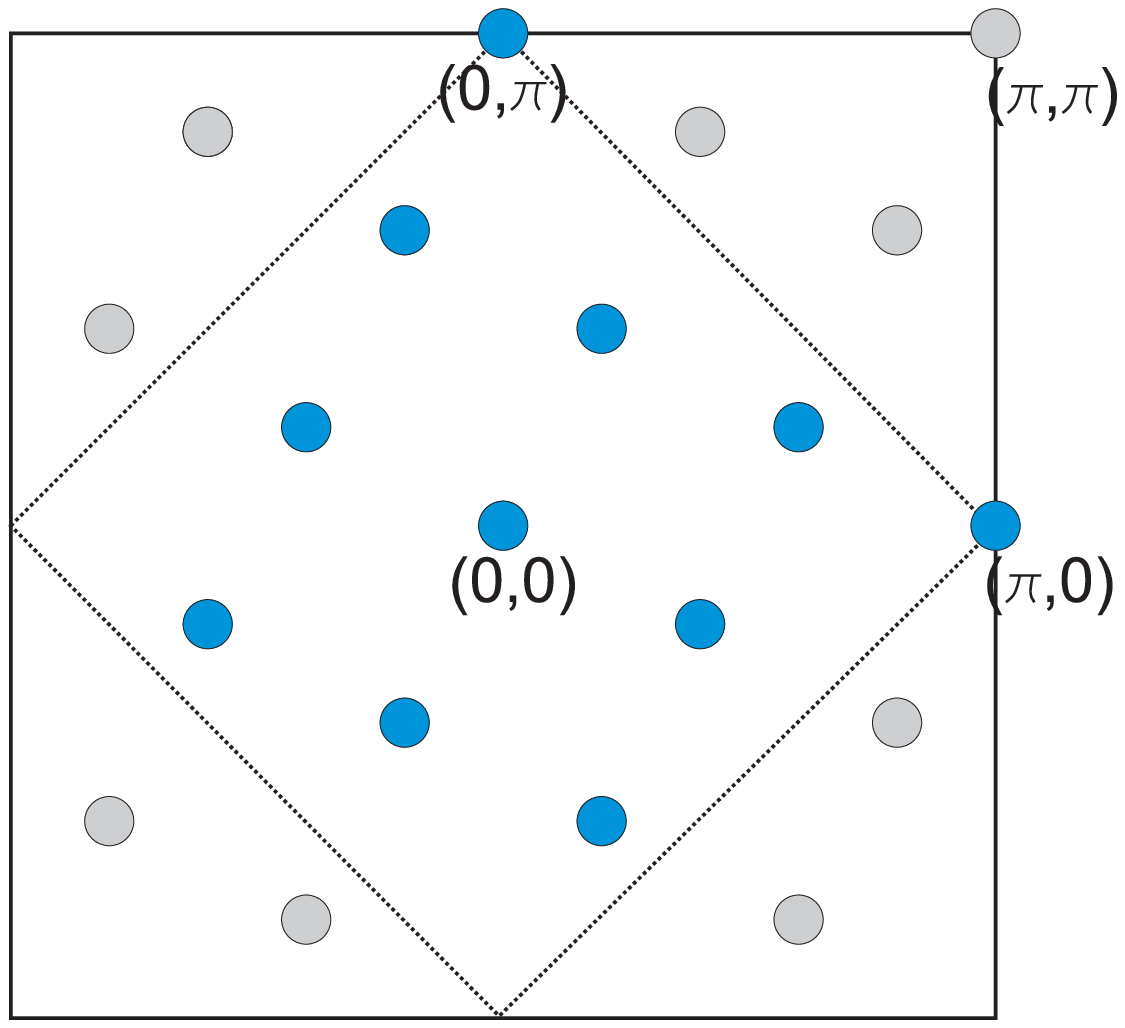}
\end{minipage} 
\begin{minipage}[c]{0.22 \textwidth}
  \vspace{0.5cm}
  c) $J=0.4~t$
  \includegraphics[width =\textwidth] {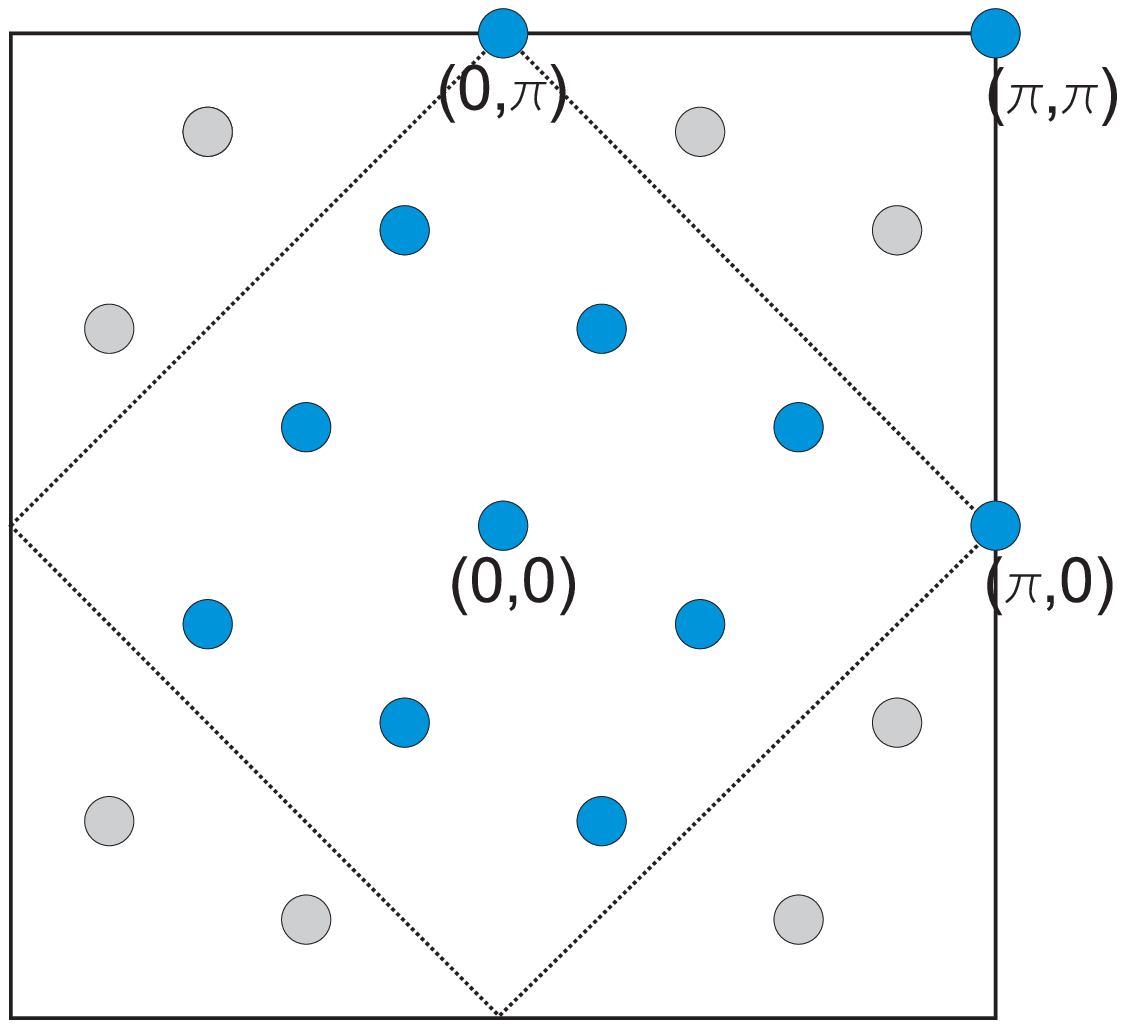}
\end{minipage}
\begin{minipage}[c]{0.22 \textwidth}
  \vspace{0.5cm}
  d) $J=0.5~t$
  \includegraphics[width =\textwidth] {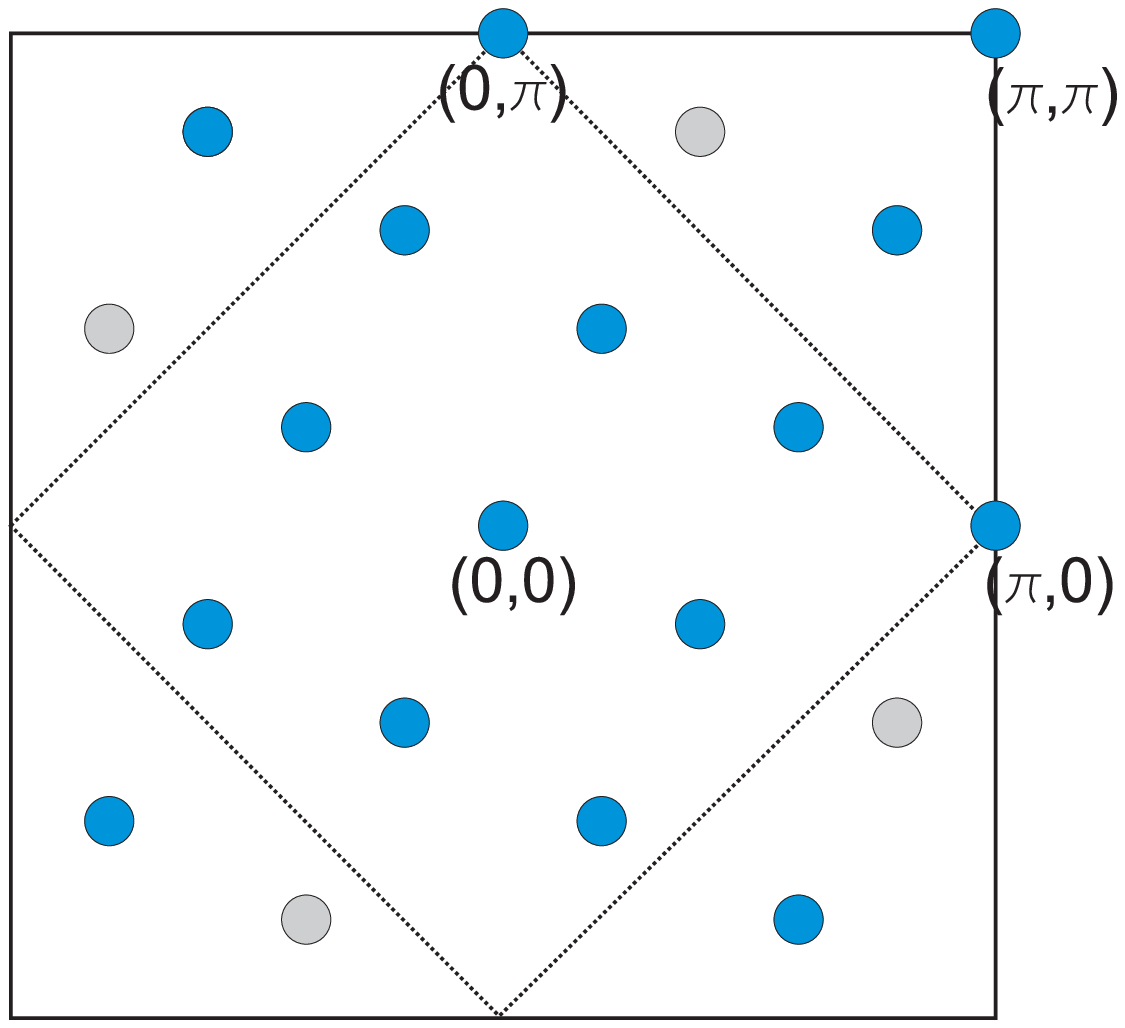}
\end{minipage}
\caption{
Luttinger volume for closed shell configuration of $N=18$
electrons on $N_0=20$ square lattice for: a) noninteracting
electrons, b) $t$-$J$ model with $0.1~t<J<0.3~t$, c) $J=0.4~t$, and d)
$J=0.5~t$. Gray (blue) circles represents (nonequivalent) wave vectors
${\bf k}$ outside (inside) the Luttinger volume.}
\end{figure}

Results for the violation of the LSR, presented in Figs.~1b-d are not
unexpected and qualitatively consistent with other (indirect)
indications of the violation of the LSR within the $t$-$J$ model,
obtained via the high-$T$ expansion of momentum distribution $n_{\bf
k}$ \cite{putt} or a straightforward interpretation of spectral
functions $A({\bf k},\omega)$, calculated on small systems
\cite{zeml}. Namely, in the range of parameters for cuprates $J \sim
0.3~t$ results within the $t$-$J$ model show too large LV, $N'>N$,
whereby the violation is modest \cite{putt,zeml}, the conclusion is
consistent with experimental finding for hole doped cuprates
\cite{shen,yosh}.

In a doped system larger $J$ plausibly enhances the antiferromagnetic
order (remaining always finite-range in small systems) which can lead
to the formation of hole pockets, which should show up in small Fermi
surface and corresponding LV forming around ${\bf
k}=(\pi/2,\pi/2)$. Indeed such tendency is evident in Figs.~1c,d where
at largest $J=0.5~t$ only the point closest to ${\bf k}=(\pi/2,\pi/2)$
remains outside the LV.

\section{Conclusions}

In conclusion, we have shown that the investigation of finite-size
systems can provide a non-trivial test of the LSR and its validity for
strongly correlated systems. Several properties of finite systems, in
particular the degeneracy of g.s., can complicate the
interpretation of results. Nevertheless, the cases with the major
violation of the LSR should be easily detectable also in small
systems. In this study we examine only prototype Hubbard
and $t$-$J$ models, still the application of the method to other
models of correlated electrons is straightforward.

It should be pointed out that in all evident (type IV) cases the
breakdown of the LSR appears by a continuous variation of GF zero
$\omega_0\sim 0$ across the chemical potential, which in a finite
system is connected with a divergence of the corresponding
$\Sigma({\bf k}, \omega \sim 0) \to \infty$. Such a transition remains
an evident possibility in an insulator even in the limit $N \to
\infty$ \cite{rosc}.  On the other hand, in a metal (where in a normal
FL the Fermi surface is determined by poles of GF) such a scenario
would represent a major modification of the LV concept and therefore
the limit $N \to \infty$ allowing for several nontrivial scenarios
should be considered with care.

For the Hubbard model on 1D chain and 2D square lattice, both for
$t'=0$ and $t' \neq 0$, we do not find a clearcut violation (type IV)
of the LSR. While this is rather a test of our approach for 1D systems
where exact results \cite{lieb,frah} indicate that for $N \to \infty$
the LV remains that of the NIE with the singularity at $k_F=\pi
N/2N_0$ away from half filling. For 2D system the validity of the LSR would
represent the confirmation of the fact that the g.s. even at large
$U>0$ is adiabatically connected to that of NIE, as far it represents
the paramagnetic state. Several examples of type II, III deviations
still point on a cautious interpretation, since systems studied
numerically are small. In particular for 2D lattice, large separation
of discrete levels for NIE in considered systems indicate that it
would be hard to observe small deviations from LSR. Another
possibility in $\textrm{D}>1$ is a change of the Luttinger-surface form while
preserving LV, allowed and expected for $U>0$. In our approach the
latter would show up with few zeroes of GF entering and leaving the
LV. Since nonequivalent zeros can hardly cross $\mu$ at the same
parameter in a finite system, this would lead to intermediate
violation of LSR. Again, our systems seem to be too small to observe
this phenomenon.

Within the $t$-$J$ model we observe the violation of the LSR,
particularly evident for the 2D systems corresponding to low doping of
the Mott insulator as relevant to superconducting cuprates. This can
be interpreted the g.s. of the model not being adiabatically connected
to the g.s. of NIE, although formally the latter would be allowed by
performing the perturbation in $J>0$ and letting $\tilde U\to \infty$
in Eq.~(\ref{eq8a}). It is not surprising that the latter
extrapolation can break the connection to NIE. In our study we follow
numerically only $J>0.1~t$. On the other hand, it is clear that $J \to
0$ case should be equivalent to the Hubbard model with $U \to \infty$.
In the latter regime, the Hubbard model (or $t$-$J$ model with $J \to
0$) close to  half-filling can show instability towards partly or
fully spin polarized states $S\gg 0$ as well as more pronounced
finite-size effects, which are only partly studied in the present
work. 

Our results are consistent with other theoretical indications for the
deviations from LSR \cite{putt,zeml} as well as experiments on
cuprates \cite{shen,yosh}. Still the speculations on the origin and
the extrapolation to a macroscopic system are delicate. While the
deviation from the LSR could indicate a general breakdown of the Fermi
liquid concept or be a sign of inherent insufficiency of the model,
the deviations could as well disappear in the limit $N \to \infty$ in
a metallic system while persisting (or not) e.g. in a Mott insulating
state.

One of the authors (P.P.) thanks M. Potthof for useful suggestions
regarding the LSR.

\end{document}